\documentclass[acus]{JAC2003}
\usepackage{graphicx}
\usepackage{epstopdf}
\usepackage{pdfsync}
\usepackage{cite}

\DeclareGraphicsExtensions{.ps,.jpg,.eps,.epsf,.EPS,.EPSF}
\newcommand{\deltamax} {\delta_{\rm max}}
\newcommand{\deltaeff} {\delta_{\rm eff}}
\newcommand{\Emax} {E_{\rm max}}

\newcommand{\Yeff} {Y_{\rm eff}}

\newcommand{\simgt} {\>\hbox{\lower0.6ex\hbox{$\sim$}\llap{\raise0.6ex\hbox{$>$}}}\>}
\newcommand{\simlt} {\>\hbox{\lower0.6ex\hbox{$\sim$}\llap{\raise0.6ex\hbox{$<$}}}\>}

\newcommand{\beq} {\begin{equation}}
\newcommand{\eeq} {\end{equation}}
\newcommand{\beqar} {\begin{eqnarray}}
\newcommand{\eeqar} {\end{eqnarray}}
\newcommand{\bseqar} {\begin{subeqnarray}}
\newcommand{\eseqar} {\end{subeqnarray}}

\begin{document}
\title{
%\vspace{-0.2 in}\hfill{\small\textmd{CBP-xxx/LBNL-xxx\quad (date)}} \\
ELECTRON CLOUD EFFECTS IN ACCELERATORS\thanks{Work supported by the US DOE under contract DE-AC02-05CH11231. 
%Invited talk presented at the ECLOUD12 Workshop (Elba, Italy, June 5-8, 2012).}
}
}
\author{M. A. Furman,\thanks{mafurman@lbl.gov} Center for Beam Physics, LBNL, Berkeley, CA 94720, \\
and CLASSE, Cornell University, Ithaca, NY 14853}
\maketitle

\begin{abstract}
We present a brief summary of various aspects of the electron-cloud effect (ECE) in accelerators. 

For further details, the reader is encouraged to refer to the proceedings of many prior workshops, either dedicated to EC or with significant EC contents, including the entire ``ECLOUD'' series \cite{CEIBA95,SantaFe1997,MBI97,HB1999,twostream00-santafe,twostream01-KEK,HB2002,ECLOUD02,SPS-scrubbing-2002,Prise2003-BNL,ECLOUD04,HB2004,HHH2004,HB2006,ECL2,ECLOUD07,ECM08,HB2008,AEC09,HB2010,ECLOUD10,CERN-GSI-mar11}. In addition, the proceedings of the various flavors of Particle Accelerator Conferences \cite{JACOW} contain a large number of EC-related publications. The ICFA Beam Dynamics Newsletter series \cite{ICFABDNL} contains one dedicated issue, and several occasional articles, on EC. An extensive reference database is the LHC website on EC \cite{EC-LHC-website}.
\end{abstract}

\maketitle

%%%%%%%%%%%%%%%%%%%%%%%%%%%%%%%%%%%%%%%%%%%%%%%%%%%%%%%%%%%%%
\section{Introduction}

The qualitative picture of the development of an electron cloud for a bunched beam is as follows:
\begin{enumerate}
\item Upon being injected into an empty chamber, a beam generates electrons by one or more mechanisms; these electrons are usually referred to as primary, or seed, electrons.
\item These primary electrons get rattled around the chamber from the passage of successive bunches.
\item As these electrons hit the chamber surface they yield secondary electrons, which are, in turn, added to the existing electron population.
\end{enumerate}

This process repeats with the passage of successive bunches. An essential ingredient of the build-up and dissipation of the EC is the secondary electron yield (SEY) of the chamber surface, characterized by the function $\delta(E)$, where $E$ is the electron-wall impact energy. The function $\delta(E)$ has a peak $\deltamax$ typically ranging in $1-4$ at an energy $E=\Emax$ typically ranging in $200-400$ eV. A convenient phenomenological parameter is the effective SEY, $\deltaeff$, defined to be the average of $\delta(E)$ over all electron-wall collisions during a relevant time window. Unfortunately, there is no simple a-priori way to determine $\deltaeff$, because it depends in a complicated way on a combination of many of the beam and chamber parameters. 

If $\deltaeff<1$, the chamber wall acts as a net absorber of electrons and the EC density $n_e$ grows linearly in time following beam injection into an empty chamber. The growth saturates when the net number of electrons generated by primary mechanisms balances the net number of electrons absorbed by the walls. 

\begin{figure*}[htb]
\centerline{\hspace{0.0in}\includegraphics[scale=0.7]{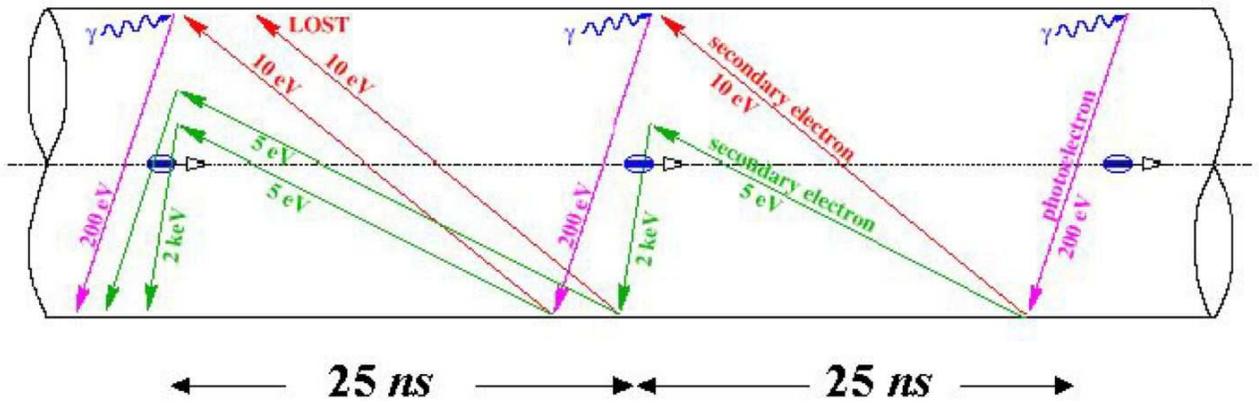}}
\caption{Cartoon illustrating the build-up of the electron cloud in the LHC for the case of 25-ns bunch spacing. The process starts with photoelectrons and is amplified by the secondary emission process. This cartoon was generated by F. Ruggiero.}
\label{fig:LHC-BIM-cartoon}
\end{figure*}

If $\deltaeff>1$, the EC initially grows exponentially. This exponential growth slows down as the space-charge fields from the electrons effectively neutralize the beam field, reducing the electron acceleration. Ultimately, the process stops when the EC space-charge fields are strong enough to repel the electrons back to the walls of the chamber upon being born, at which point $\deltaeff$ becomes $=1$. At this point, the EC distribution reaches a dynamical equilibrium characterized by rapid temporal and spatial fluctuations, determined by the bunch size and other variables. For typical present-day storage rings, whether using positron or proton beams, the average $n_e$ reaches a level $\sim10^{10-12}$ m$^{-3}$, the energy spectrum of the electrons typically peaks at an energy below $\sim100$ eV, and has a high-energy tail reaching out to keV's. In more detail, however, the EC distribution reaches a dynamical equilibrium characterized by temporal and spatial fluctuations. The temporal fluctuations span a typical range $10^{-12}-10^{-6}$ s, depending on the bunch length and intensity, and on the bunch train length and fill pattern. Spatial fluctuations typically span the range $10^{-9}-10^{-2}$ m, depending on the transverse bunch size and transverse dimensions of the vacuum chamber, and external magnetic field if any. The density $n_e$ gradually decays following beam extraction, or during the passage of a gap in the beam. The decay rate is controlled by the low-$E$ value (typically $E\simlt20$ eV) of $\delta(E)$. In general, there is no simple, direct correlation between the rise time and the fall time of the buildup of $n_e$ \cite{furman-PAC03-form-dissip}. Figure~\ref{fig:LHC-BIM-cartoon} illustrates the build-up of the electron cloud in the LHC.

%%%%%%%%%%%%%%%%%%%%%%%%%%%%%%%%%%%%%%%%%%%%

The ECE combines many parameters of a storage ring such as bunch intensity, size and spacing, beam energy \cite{rumolo-PRL100-2008}, vacuum chamber geometry, vacuum pressure, and electronic properties of the chamber surface material such as photon reflectivity $R_\gamma$, effective photoelectric yield (or quantum efficiency) $\Yeff$, the SEY, the secondary emission spectrum \cite{furman-HB2006,cimino-PRL93-2004}, etc.  

In regions of the storage ring with an external magnetic field, such as dipole bending magnets, quadrupoles, etc., the EC distribution develops characteristic geometrical patterns. For typical magnetic fields in the range $B=0.01-5$ T and typical EC energies $<100$ eV, the electrons move in tightly-wound spiral trajectories about the field lines. Thus in practice, in a bending dipole, the electrons are free to move in the vertical ($y$) direction, but are essentially frozen in the horizontal ($x$). As a result, the $y$-kick imparted by the beam on a given electron has an $x$ dependence that is remembered by the electron for many bunch passages. It often happens that the electron-wall impact energy equals $\Emax$ at an $x$-location less than the horizontal chamber radius. At this location $\delta(E)=\deltamax$, hence $n_e$ is maximum, leading to characteristic high-density vertical stripes symmetrically located about $x=0$ \cite{zimm-ECLOUD02-large-effect}. For quadrupole magnets, the EC distribution develops a characteristic four-fold pattern, with characteristic four-fold stripes \cite{arduini-EPAC04}.

In summary, the electron-cloud formation and dissipation:

\begin{itemize}
\item Is characterized by rich physics, involving many ingredients pertaining to the beam and its environment.
\item Involves a broad range of energy and time scales. 
\item Is always undesirable in particle accelerators.
\item Is often a performance-limiting problem, especially in present and future high-intensity storage rings.
\item Is challenging to accurately quantify, predict and extrapolate.
\end{itemize}

The electron cloud has been shown to be detrimental to the performance of many storage rings, and is a concern for future such machines, which typically call for high beam intensity and compact vacuum chambers. At any given storage ring, adverse effects may include one or more of the following: sudden, large, vacuum pressure rise; beam instabilities; emittance growth; interference with diagnostic instrumentation; excessive heat deposition on the chamber walls; etc. Mitigation mechanisms have been required in most cases in order to reach, or exceed, the design performance of the machine. 

A more extensive summary of the ECE and its history is presented in Ref.~\cite{furman-buildup-ECLOUD10}.

%%%%%%%%%%%%%%%%%%%%%%%%%%%%%%%%%%%%%%%%%%%%%%%%%%%%%%%%%%%%%%%%%%%%%%%
\section{Primary and secondary electrons}

The main sources of primary electrons are: photoemission from synchrotron-radiated photons striking the chamber walls; ionization of residual gas; and electron generation from stray beam particles striking the walls of the chamber. Depending on the type of machine, one of these three processes is typically dominant. For example, in positron or electron storage rings, upon traversing the bending magnets, the beam usually emits copious synchrotron radiation with a $\sim$keV critical energy, yielding photoelectrons upon striking the vacuum chamber. In proton rings, the process is typically initiated by ionization of residual gas, or from electron generation when stray beam particles strike the chamber. A notable exception is the LHC, which is the first proton storage ring ever built in which the beam emits significant synchrotron radiation, $\sim0.4$ photons per proton per bending magnet traversal, with a photon critical energy $\sim 44$ eV \cite{zimm-LHCPR95}. In this case, photoemission is the dominant primary mechanism.

Primary emission mechanisms are usually insufficient to lead to a significant EC density. However, the average electron-wall impact energy is typically $\sim$100--200 eV, at which the SEY function $\delta(E)$ is significant. If the effective SEY is $>1$, secondary emission readily exponentiates in time, which can lead to a large amplification factor, typically a few orders of magnitude, over the primary electron density, and to strong temporal and spatial fluctuations in the electron distribution \cite{rumolo-PRSTAB-012801-2001}. This compounding effect of secondary emission is usually the main determinant of the strength of the ECEs, and is particularly strong in positively-charged bunched beams (in negatively-charged beams, the electrons born at the walls are pushed back towards the walls with relatively low energy, typically resulting in relatively inefficient secondary emission).

Photoemission and secondary electron emission depend differently on the beam properties: photoelectron emission behaves linearly in beam intensity, is very sensitive to beam energy, and is independent of the sign of the beam particle charge, while secondary emission behaves nonlinearly in beam intensity, is not very sensitive to beam energy, and is sensitive to the sign of the beam particle charge. These features allow, in principle, to disentangle the effects of primary from secondary electrons, given sufficient flexibility in the machine operation as in CESRTA (see below).

%%%%%%%%%%%%%%%%%%%%%%%%%%%%%%%%%%%%%%%%%%%%%%%%%%%%%%%%%%%%%
\section{Conditioning and Mitigation}

Storage ring vacuum chambers are fabricated of ''technical metals.'' Such materials have rough surfaces and contain impurities, typically concentrated at the surface. For such surfaces, the SEY gradually decreases in time with machine operation owing to the bombardment of the very electrons in the cloud. Such ``conditioning effect'' has been consistently observed in storage rings, and is of course beneficial to the performance of the machine. Typically, it is observed that $\deltamax$ decreases rapidly (typically hours to days) upon machine operation startup, and then effectively reaches a limit. Indeed, as $\deltamax$ decreases, the EC intensity decreases, leading to a diminished electron-wall bombardment, hence to a slower conditioning rate. This exponential slowing down, in effect, sets a practical limit on the lowest value of $\deltamax$ that is achievable via this phenomenon. Recent experience at the SPS and LHC \cite{jimenez-rumolo-cham2012} is consistent with prior experience at many other machines, namely that $\deltamax$ decreases rapidly but does not go far enough to avoid all EC detrimental effects.

Even if $\deltamax$ were to decrease via the conditioning effect to its natural limit \cite{cimino-PRL-109-064801-2012,larciprete-PRSTAB16-011002}, it is not guaranteed to be low enough to avoid undesirable ECE's. For this reason, deliberate mitigation mechanisms are typically implemented in present-day and future storage rings. Mitigation mechanisms can be classified into passive and active. Passive mechanisms that have been employed at various machines include:

\begin{itemize}
\item Coating the chamber with low-emission substances such as TiN \cite{kennedy-PAC97,he-PAC01-SNS-TiN-coating}, TiZrV \cite{fischer-EC-at-RHIC-PRSTAB11-2008,hseuh-PAC05-RHIC-RT-coatings,suetsugu-coatings-1-NIMPR2005,suetsugu-coatings-2-NIMPR2006,suetsugu-coatings-PAC07,suetsugu-vac-system-BDNL48,AEC09,LHC-DR-RT-coatings,AEC09,LHC-DR-RT-coatings} and amorphous carbon (a-C) \cite{shaposhnikova-carbon-coat-PAC09,yin_vallgren-carbon-coat-IPAC10}.
\item Etching grooves on the chamber surface in order to make it effectively rougher, thereby decreasing the effective quantum efficiency via transverse grooves \cite{baglin-candidate-EPAC98} or the effective SEY via longitudinal grooves \cite{stupakov-ECLOUD04-grooves,suetsugu-grooves-IPAC10}.
\item Implementing weak solenoidal fields ($\sim$10--20 G) to trap the electrons close to the chamber walls, thus minimizing their detrimental effects on the beam \cite{kulikov-PAC01-TPPH100,funakoshi-PAC01-RPPH131}
\end{itemize}

In terms of active mechanisms, clearing electrodes \cite{suetsugu-clearing-electrode-PAC09,suetsugu-clearing-electrode-IPAC10} show significant promise in controlling the electron cloud development. If an electron cloud is unavoidable and problematic, active mechanisms that have been employed to control the stability of the beam include tailoring the bunch fill pattern \cite{decker-PAC01-complicated} and increasing the storage ring chromaticity \cite{rumolo-PRSTAB-012801-2001}. Fast, single-bunch, feedback systems are under active investigation as an effective mechanism to stabilize electron-cloud induced coherent instabilities \cite{vay-IPAC10-SPS-FDBK,fox-IPAC10-SPS-FDBK}. 

%%%%%%%%%%%%%%%%%%%%%%%%%%%%%%%%%%%%%%%%%%%%%%%%%%%%%%%%%%%%%
\section{Simulation of the ECE}
Broadly speaking, depending on the approximations implemented, EC simulation codes in use today are of three kinds:

\begin{itemize}
\item Build-up codes.
\item Instability codes.
\item Self-consistent codes.
\end{itemize}

Build-up codes make the approximation that the beam is a prescribed function of space and time, and therefore is nondynamical. The electrons, on the other hand, are fully dynamical. With this kind of code one can study the build-up and decay of the EC, its density distribution, and its time and energy scales, but not the effects of the EC on the beam\footnote{Actually, these codes do allow the computation of the dipole wake induced by the EC on the beam, which in turn allows a first-order computation of the coherent tune shift of successive bunches of the beam.}. These codes may include a detailed model of the electron-wall interaction, and come in 2D and 3D versions. 2D codes are well suited to study the EC in certain isolated regions of a storage ring, such as in the body of magnets, and field-free regions. 3D codes are used to study the EC in magnetic regions that are essentially 3D in nature, such as fringe fields and wigglers.

Instability codes aim at studying the effects on the beam by an initially prescribed EC. In these codes the beam particles are fully dynamical, while the dynamics of the cloud electrons is limited. For example, the electron-wall interaction may be simplified or non-existent, and/or the electron distribution may be refreshed to its initial state with the passage of successive bunches.

Self-consistent codes aim to study the dynamics of the beam and the electrons under their simultaneous, mutual, interaction. Such codes are far more computationally expensive than either of the above-mentioned ``first-order'' codes, and represent the ultimate logical stage of the above-mentioned simulation code efforts. 

In many cases of interest, the net electron motion in the longitudinal direction, i.e. along the beam direction, is not significant, hence the electron cloud is sensibly localized. For this reason, in first approximation, it makes sense to study it at various locations around the ring independently of the others. In addition, given that the essential dynamics of the electrons is in the transverse plane, i.e. perpendicular to the beam direction, two-dimensional simulations are also a good first approximation to describe the build-up and decay. In some cases, such as the PSR, electron generation, trapping and ejection from the edges of quadrupole magnets is now known to be significant, and these electrons act as seeds for the EC buildup in nearby drift regions \cite{macek-ECLOUD10-quad-trapp}.

A comprehensive online repository containing code descriptions and contact persons has been developed by the CARE program \cite{HHH-ecloud-code-website}. 

\begin{figure}[htb]
\centerline{\hspace{0.0in}\includegraphics[scale=0.6]{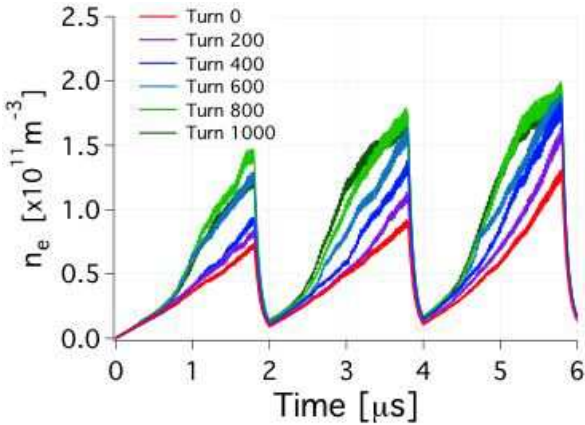}}
\caption{The EC density $n_e$ as a function of time over a 6-$\mu$s time window, showing the passage of a 3-batch beam (the revolution period is $\sim$23 $\mu$s). Each trace represents the evolution after the number of revolutions indicated. For example, the blue trace (``turn 400'') shows the window after 399 turns have elapsed. The red trace (``turn 0'') shows the evolution of $n_e$ at beam injection; this trace is in excellent agreement with the result of a build-up code, as it should, in which the proton distribution is kept frozen at its initial state (a 3D gaussian distribution). In this exercise, the electron distribution was reinitialized at every turn at the beginning of the train. Thus the fact that the ``turn 1000'' trace is a factor $\sim2$ times larger than the ``turn 0'' trace is attributable only to the evolution of the proton distribution in the beam after 1000 turns.}
\label{fig:vay-IPAC12-TUEPPB006}
\end{figure}

Self-consistent codes are beginning to yield useful results. We present here one such example obtained with the code WARP/POSINST, pertaining to the SPS \cite{vay-IPAC12-TUEPPB006}. In this case, a train of three beam batches, each consisting of 72 bunches, was simulated using a massively parallel computer at NERSC. The goal of the simulation was primarily to assess the impact of the evolution of the proton distribution in the beam on the EC density, as compared to the EC density evolution produced by a build-up code, in which the proton distribution is frozen in time. Fig.~\ref{fig:vay-IPAC12-TUEPPB006} shows some of the results of this exercise. The conclusion is that, after 1000 turns, the actual proton distribution leads to a 50--100\% increase in the estimate of $n_e$ relative to the case in which the proton distribution is kept frozen at its initial state. While this result is suggestive, it must still be considered preliminary because of the approximations employed, notably that of a constant focusing lattice and the fact that the EC distribution was reinitialized at avery turn (a fully self-consistent simulation, in which both the EC and the proton distributions evolve in time in response to each other has also been carried out \cite{vay-IPAC12-TUEPPB006}).

%%%%%%%%%%%%%%%%%%%%%%%%%%%%%%%%%%%%%%%%%%%%%%%%%%%%%%%%%%%%%%%%%%%%%%%%%%%%%%%%%%
\section{The CESRTA program}
A significant, dedicated systematic R\&D program to understand the EC and low-emittance tuning has been ongoing at Cornell University for $\sim5$ years based on the CESR storage ring. The e$^+$e$^-$ collider CESR was decommissioned and the CLEO detector removed. Wigglers were added to the storage ring, along with an extensive array of diagnostic instrumentation intended to analyze the EC. This revamped storage ring (the CESR Test Accelerator, or CESRTA) is intended as a prototype for the damping rings of a possible future e$^+$e$^-$ linear collider \cite{palmer-PAC09-cesrta-conversion}. A major report will describe the R\&D effort in detail \cite{cesrta-phase-1-report}.

As a test accelerator, CESRTA has unprecedented operational flexibility, specifically:

\begin{itemize}
\item Essentially all beam time is devoted to machine studies.
\item The injector allows for an almost arbitrary fill pattern.
\item The beam species is selectable (e$^+$ or e$^-$), although the two species move in opposite directions in the beam pipe.
\item The beam energy is tunable within the range $\sim2-5$ GeV
\item The bunch intensity is selectable.
\end{itemize}

The new diagnostic devices include: retarding-field analyzers (RFA's) at many locations, magnetized or not; shielded pick-ups (SPU's); a microwave transmission setup; filtered and gated beam position monitors (BPM's); etc. In addition, an array of special-purpose devices have been installed including: an in-situ SEY measuring device; a low-magnetic-field chicane, transplanted from PEP-II at SLAC; various sections of beam pipe with low-emission coatings or grooved surfaces; and clearing electrodes. RFA's allow the measurement of the spatially-resolved, time-averaged, electron flux at the walls of the chamber. The SPU's allow the measurement of the electron flux at the walls of the chamber with a time resolution of $\sim1$ ns. The BPM's, by themselves or in combination with a beam pinger and a feedback damping system, allow the measurement of bunch-by-bunch frequency spectra and coherent tunes. x-ray beam-size monitors allow the measurement of beam size bunch-by-bunch and turn-by-turn.

As part of the CESRTA R\&D, a broad-based program of developing, comparing and benchmarking electron cloud buildup simulation codes, and to a much lesser extent beam dynamics codes, was initiated in 2008 and continues today. Specifically CESRTA input parameters have been used as input to the simulation codes ECLOUD~\cite{ECLOUD04:143to152,PRSTAB5:121002}, 
CLOUDLAND~\cite{PRSTAB5:124402,KEK:INT2003:2}, POSINST~\cite{furman-MBI97-PEPII,furman-pivi-prob-SEY-PRSTAB02}, 
WARP/POSINST~\cite{PAC09:FR5RFP078} and PEHTS~\cite{PAC01:TPPH096}, and the results compared against measurements. By iterating this process, EC-related parameters that are not well known were pinned down, allowing more reliable extrapolations to the future ILC damping rings. The main parameters that are not well known are those pertaining to the electronic surface properties, i.e. photon reflectivity; photoemission yield or quantum efficiency (QE); photoemission spectrum; and secondary electron yield and spectrum \cite{dugan-crittenden-these-proc}.

In addition, a new photon-tracking code, SYNRAD3D \cite{synrad3d}, has been developed and implemented, which allows the tracking of synchrotron radiation emitted by the beam as it traverses magnetic elements. The code allows for the description of the actual beam size at the emission point, as well as the actual description of the vacuum chamber geometry and external magnetic fields for the entire ring. Models for the photon reflectivity and quantum efficiency have been incorporated. The outcome of this code is the photoelectron emission distribution along the perimeter of the chamber cross section at any desired point in the ring. This photoelectron distribution is fed as an input to the above-mentioned build-up codes. A simpler code of this nature was developed earlier in the context of the LHC EC effort \cite{zimm-LHC-PN237}. 

\begin{figure}[htb]
\centerline{\hspace{0.0in}\includegraphics[scale=0.23]{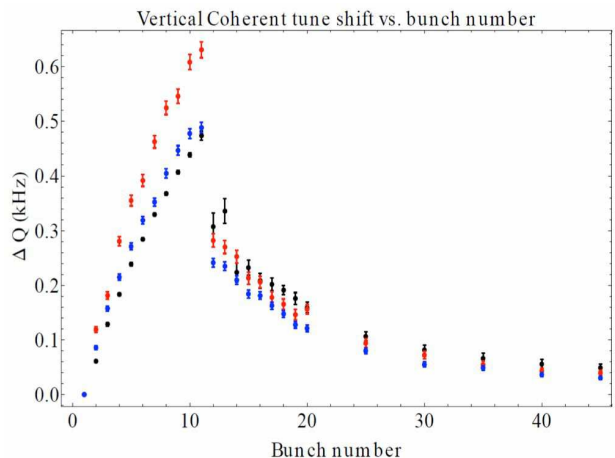}}
\caption{Measured tune shifts (black points) vs. bunch number, for a train of ten 0.75-mA/bunch, 5.3 GeV, positron bunches with 14 ns spacing, followed by witness bunches \cite{dugan-IPAC12-WEYA02}. Red points are computed (using POSINST) based on a simplified assumption for the incident photon distribution consisting of a direct component plus a uniform background (free parameter) of scattered photons. Blue points are computed using results for the photoelectron emission distribution obtained from SYNRAD3D (with no free parameters for the radiation) as input to POSINST. The good agreement between measurements and simulations gives confidence in the EC model implemented in the code POSINST. The computation employing the SYNRAD3D results is clearly in better agreement with the measurements than that using a simplified photoemission distribution model.}
\label{fig:cesrta-tunes}
\end{figure}

By adjusting the bunch train length and adding a ``witness bunch'' at various distances after the end of the train, one is able to disentangle the effects of the photoelectrons from the secondary electrons. A comparison of a simulation vs. measurements at CESRTA is shown in Fig.~\ref{fig:cesrta-tunes} \cite{dugan-IPAC12-WEYA02}.

%%%%%%%%%%%%%%%%%%%%%%%%%%%%%%%%%%%%%%%%%%%%%%%%%%%%%%%%%%%%%%%
\section{Conclusions}
\begin{itemize}
\item The ECE is an ubiquitous phenomenon for intense beams. The phenomenon spans a broad range of charged-particle storage rings. 
\item The ECE is important inasmuch as it limits machine performance, especially for high-intensity future machines. 
\item The ECE is interesting, as it involves in an essential way various areas of physics, such as: surface geometry and surface electronics; beam intensity and particle distribution; beam energy; residual vacuum pressure in the chamber; certain magnetic features of the storage ring; and other areas.
\item Simulation codes are getting better and better in their detailed modeling capabilities and predictive ability.
\item Enormous progress has been made since 1995, with a disproportionate credit due to CESRTA and CERN over the past few years. Better and more refined electron detection mechanisms are now deployed. Simulation codes are getting better and better calibrated against measurements.
\item Phenomelogical rules of thumb are appearing that tell us the conditions under which the ECE is serious, but not (yet) the conditions under which itÕs guaranteed to be safe.
\end{itemize}

%%%%%%%%%%%%%%%%%%%%%%%%%%%%%%%%%%%%%%%%%%%%%%%%%%%%%%%%%%%%%%%
\section{Epilogue}
This workshop is dedicated to the memory of Francesco Ruggiero (1957-2007). I met Francesco on many occasions during my career. I feel honored to have met him and grateful for what I learned from him. I am especially grateful to Francesco for his strong support of electron-cloud R\&D effort at CERN and elsewhere. The knowledge that has come out of this program, plus the recent experience at the LHC and SPS, have already greatly benefitted the field as a whole, and will continue to benefit the design and reliability of accelerators worldwide for a long time to come. This workshop is rightfully dedicated to Francesco's memory. 

%%%%%%%%%%%%%%%%%%%%%%%%%%%%%%%%%%%%%%%%%%%%%%%%%%%%%%%%%%
\section{acknowledgments}
Over the years I have greatly benefitted from discussions and/or collaboration with many colleagues at ANL, BNL, CERN, Cornell, FNAL, Frascati, KEK, LANL, LBNL, SLAC and TechX---I am grateful to all of them, too numerous to list here. I want to express my special thanks to Roberto Cimino and Frank Zimmermann for organizing this productive and enlightening workshop. We are grateful to NERSC for supercomputer support.

%%%%%%%%%%%%%%%%%%%%%%%%%%%%%%%%%%%%%%%%%%%%%%%%%%%%%%%%%%

%\section*{DISCLAIMER}
%
%This document was prepared as an account of work sponsored by the United States Government. While this document is believed to contain correct information, neither the United States Government nor any agency thereof, nor The Regents of the University of California, nor any of their employees, makes any warranty, express or implied, or assumes any legal responsibility for the accuracy, completeness, or usefulness of any information, apparatus, product, or process disclosed, or represents that its use would not infringe privately owned rights. Reference herein to any specific commercial product, process, or service by its trade name, trademark, manufacturer, or otherwise, does not necessarily constitute or imply its endorsement, recommendation, or favoring by the United States Government or any agency thereof, or The Regents of the University of California. The views and opinions of authors expressed herein do not necessarily state or reflect those of the United States Government or any agency thereof, or The Regents of the University of California.
%
%Ernest Orlando Lawrence Berkeley National Laboratory is an equal opportunity employer.

\end{document}